\newif\ifprintfig
\printfigtrue
\pdfoutput=1

\documentclass{emulateapj}
\usepackage{graphicx}
\usepackage{subfigure}
\usepackage{amsmath}

\newcommand\clock{\count0=\time \divide\count0 by 60
    \count1=\count0 \multiply\count1 by -60 \advance\count1 by \time
    \number\count0:\ifnum\count1<10{0\number\count1}\else\number\count1\fi}

\shortauthors{Beerman et al}
\shorttitle{Partially Resolved Clusters}

\begin{document}

\title{{The Panchromatic Hubble Andromeda Treasury III.  Measuring Ages and Masses of Partially Resolved Stellar Clusters} \footnote{Based on observations made with the NASA/ESA Hubble Space Telescope, obtained from the Data Archive at the Space Telescope Science Institute, which is operated by the Association of Universities for Research in Astronomy, Inc., under NASA contract NAS 5-26555.}}

\author{Lori C. Beerman\altaffilmark{1}, L.\ Clifton Johnson\altaffilmark{1}, Morgan Fouesneau\altaffilmark{1}, Julianne J.\ Dalcanton\altaffilmark{1}, Daniel R.\ Weisz\altaffilmark{1}, Anil C.\ Seth\altaffilmark{2}, Ben F.\ Williams\altaffilmark{1}, Eric F. Bell\altaffilmark{3}, Luciana C. Bianchi\altaffilmark{4}, Nelson Caldwell\altaffilmark{5}, Andrew E. Dolphin\altaffilmark{6},  Dimitrios A. Gouliermis\altaffilmark{7,8}, Jason S. Kalirai\altaffilmark{9}, S{\o}ren S. Larsen\altaffilmark{10}, Jason L. Melbourne\altaffilmark{11}, Hans-Walter Rix\altaffilmark{8}, and Evan D. Skillman\altaffilmark{12}}

\email{beermalc@astro.washington.edu}
\altaffiltext{1}{Department of Astronomy, University of Washington, Box 351580, Seattle, WA 98195, USA}
\altaffiltext{2}{Department of Physics and Astronomy, University of Utah, Salt Lake City, UT 84112, USA}
\altaffiltext{3}{Department of Astronomy, University of Michigan, 500 Church Street, Ann Arbor, MI 48109, USA}
\altaffiltext{4}{Department of Physics and Astronomy, Johns Hopkins University, 3400 North Charles Street, Baltimore, MD 21218, USA}
\altaffiltext{5}{Harvard-Smithsonian Center for Astrophysics, 60 Garden Street, Cambridge, MA 02138, USA}
\altaffiltext{6}{Raytheon Company, 1151 East Hermans Road, Tucson, AZ 85756, USA}
\altaffiltext{7}{Institut f\"ur Theoretische Astrophysik, Zentrum f\"ur Astronomie der Universit\"at Heidelberg, Albert-Ueberle-Stra{\ss}e~2, 69120 Heidelberg, Germany}
\altaffiltext{8}{Max-Planck-Institut f\"ur Astronomie, K\"onigstuhl 17, 69117 Heidelberg, Germany}
\altaffiltext{9}{Space Telescope Science Institute, 3700 San Martin Drive, Baltimore, MD 21218, USA}
\altaffiltext{10}{Department of Astrophysics, IMAPP, Radboud University Nijmegen, P.O. Box 9010, 6500 GL Nijmegen, The Netherlands}
\altaffiltext{11}{Caltech Optical Observatories, Division of Physics, Mathematics and Astronomy, Mail Stop 301-17, California Institute of Technology, Pasadena, CA 91125, USA}
\altaffiltext{12}{Department of Astronomy, University of Minnesota, 116 Church Street SE, Minneapolis, MN 55455, USA}

\begin{abstract}

The apparent age and mass of a stellar cluster can be strongly affected by stochastic sampling of the stellar initial mass function, when inferred from the integrated color of low mass clusters ($\lesssim$ $10^4$ $M_{\odot}$).  We use simulated star clusters to show that these effects are minimized when the brightest, rapidly evolving stars in a cluster can be resolved, and the light of the fainter, more numerous unresolved stars can be analyzed separately.  When comparing the light from the less luminous cluster members to models of unresolved light, more accurate age estimates can be obtained than when analyzing the integrated light from the entire cluster under the assumption that the initial mass function is fully populated.  We show the success of this technique first using simulated clusters, and then with a stellar cluster in M31.  This method represents one way of accounting for the discrete, stochastic sampling of the stellar initial mass function in less massive clusters and can be leveraged in studies of clusters throughout the Local Group and other nearby galaxies.

\end{abstract}

\vfill
\clearpage

\section{Introduction}  \label{sec:intro}

Stellar clusters are used as tools in the study of star formation, stellar evolution, and galactic evolution because they contain numerous stars of the same age, distance, and metallicity.  Their scientific utility rests on our ability to accurately measure their ages, masses, and metallicities.

Resolved observations of individual stars provide the most direct way of measuring the properties of clusters.  By resolving individual stars, one can analyze the resulting color-magnitude diagrams (CMDs) by comparing them to isochrones, allowing one to derive the age, extinction, metallicity, and distance of an individual cluster \citep[e.g.,][]{Hodge83, Elson88, Piatti07}.  An additional advantage is that one can obtain an estimate of extinction that is independent of model colors.  However, this approach is limited by the angular resolution of the available images.  If a cluster is too distant, there will not be enough individual stars resolved to use isochrone fitting.  The distance limit for analyzing clusters through resolved stars varies with the surface brightness of the cluster and the resolution of the telescope, but in practice isochrone fitting has rarely been used beyond a few Mpc, even with the Hubble Space Telescope (HST). 

If the stars in a cluster are unresolved, then the properties of the cluster can only be derived from the integrated light of the entire cluster \citep[e.g.,][]{Larsen09}.  Such studies have less stringent requirements for angular resolution, and can be carried out at larger distances (up to $\sim$ 50 Mpc; \citealt{Adamo10}).  Thus, analysis of the integrated light of unresolved star clusters will remain the method of choice for deriving ages and masses of clusters in more distant galaxies, which span the largest possible range of environments and star formation modes.

There is growing awareness of the potential limitations of unresolved cluster studies.  One of the most severe is stochastic sampling of the stellar initial mass function (IMF), which strongly affects the integrated colors and fluxes of low mass clusters.  Discrete sampling of the IMF affects the present day mass function, altering a cluster's color and luminosity from what would be expected if the IMF were continuously populated up to the highest possible stellar mass, and causes potentially rapid changes in color and magnitude throughout the lives of the cluster \citep{Piskunov11}.  Clusters with masses less than a few $10^4$ $M_{\odot}$ are subject to the effects of small number statistics at the upper end of their mass function, and thus their light can be dominated by just a few post-main sequence stars.  A small variation in their number or evolutionary state can cause two otherwise similar clusters to have drastically different integrated colors (up to five magnitudes) \citep[e.g.,][see also Figure~\ref{fig:example} in \S\ref{sec:problem}]{Barbaro77, Girardi93, Girardi95, Santos97, Brocato99, Lancon00, Bruzual02, Anders04, Raimondo05, Cervino06, Deveikis08, Popescu10a}.  This spread in colors leads to large systematic errors when deriving the clusters' properties.  Stochastic effects have been shown to be the largest source of error when deriving masses and ages for unresolved clusters  \citep[e.g.,][]{Lancon00, Cervino04, Apellaniz09, Piskunov09, Popescu10b, Fouesneau10}.

New techniques have been developed to more accurately age-date low mass clusters, by taking stochastic sampling into account.  These improvements are imperative, given that current surveys are delving more deeply into the low mass regime.  Specifically, recent studies such as \citet{Deveikis08}, \citet{Fouesneau10}, and \citet{Popescu10a, Popescu10b} have moved away from the traditional population synthesis models towards discrete population models, where the mass function is explicitly considered as a discrete distribution of stars.  The main ingredients of such techniques are large collections of discretely sampled synthetic clusters from Monte-Carlo simulations \citep[e.g.,][]{daSilva12}, which can be compared with the observed photometry.  Alternatively, individual filters can be down-weighted to reflect the degree of stochasticity \citep{Apellaniz09}.

The growing body of literature on stochastic fluctuations of clusters has clearly shown that these fluctuations are due to the random presence of massive upper main sequence and evolving stars within a population.  There is also evidence that the flux from the main sequence is fairly stable in comparison (within a few tens of percent; \citealt{Lancon08}).  As the quality of the observations has improved, we are currently able to resolve the most massive cluster stars out to distances of about 4-5 Mpc.  In many cases, one can explicitly identify the tiny subpopulation of rapidly evolving luminous stars, and exclude this small stochastically sampled population from analysis of the integrated light.  Removing the post-main sequence stars from the analysis reduces the effect of stochasticity, offering a promising way forward for deriving accurate ages of low mass clusters.  

In this paper, we explore a method for analyzing only the unresolved part of a young cluster's light, which in the ideal case includes only stars up through the main sequence turnoff.  This approach minimizes some of the troublesome effects related to the stochastically sampled upper end of a cluster's stellar mass function.  Since younger clusters are more susceptible to stochastic fluctuations and are found in abundance in nearby galaxies, we focus on minimizing errors for the younger clusters.  We demonstrate the effectiveness of this method using synthetic clusters.  We then apply this method to a cluster in M31 from the Panchromatic Hubble Andromeda Treasury (PHAT; \citealt{Dalcanton12}).

This paper is organized as follows:  In \S\ref{sec:problem} we demonstrate the problem of stochastic fluctuations in the integrated colors of clusters.  In \S\ref{sec:method} we describe our method of using only unresolved light to study partially resolved clusters.  We show the results of a simultaneous derivation of age, mass, and extinction for simulated clusters in \S\ref{sec:tests}.  In \S\ref{sec:application} we test our method on a cluster in the PHAT survey which has independent measurements of its parameters.  We discuss practical aspects of using this method in \S\ref{sec:discussion} and state our conclusions in \S\ref{sec:conclusion}.

\section{Stochastic Fluctuations}  \label{sec:problem}

A cluster's light is dominated by the brightest, most massive stars.  A small stochastic difference in the number or evolutionary state of these stars can cause two otherwise similar clusters to have drastically different integrated colors.  This effect is very pronounced in young clusters ($<$  100 Myr), whose flux may be dominated by a small number of bright red giants or supergiants that bias the integrated color towards the red.  To illustrate the effects of stochasticity in a young cluster, we created a synthetic example that shows the dramatic impact that one evolved star may have on a cluster's integrated light, similar to the examples shown in \citet{Santos97}.  For consistency with the PHAT survey, we will consider filters used in this survey, in particular F336W, F475W, F814W, and F160W.

\begin{figure}[h]
\centerline{
\includegraphics[width=0.52\textwidth]{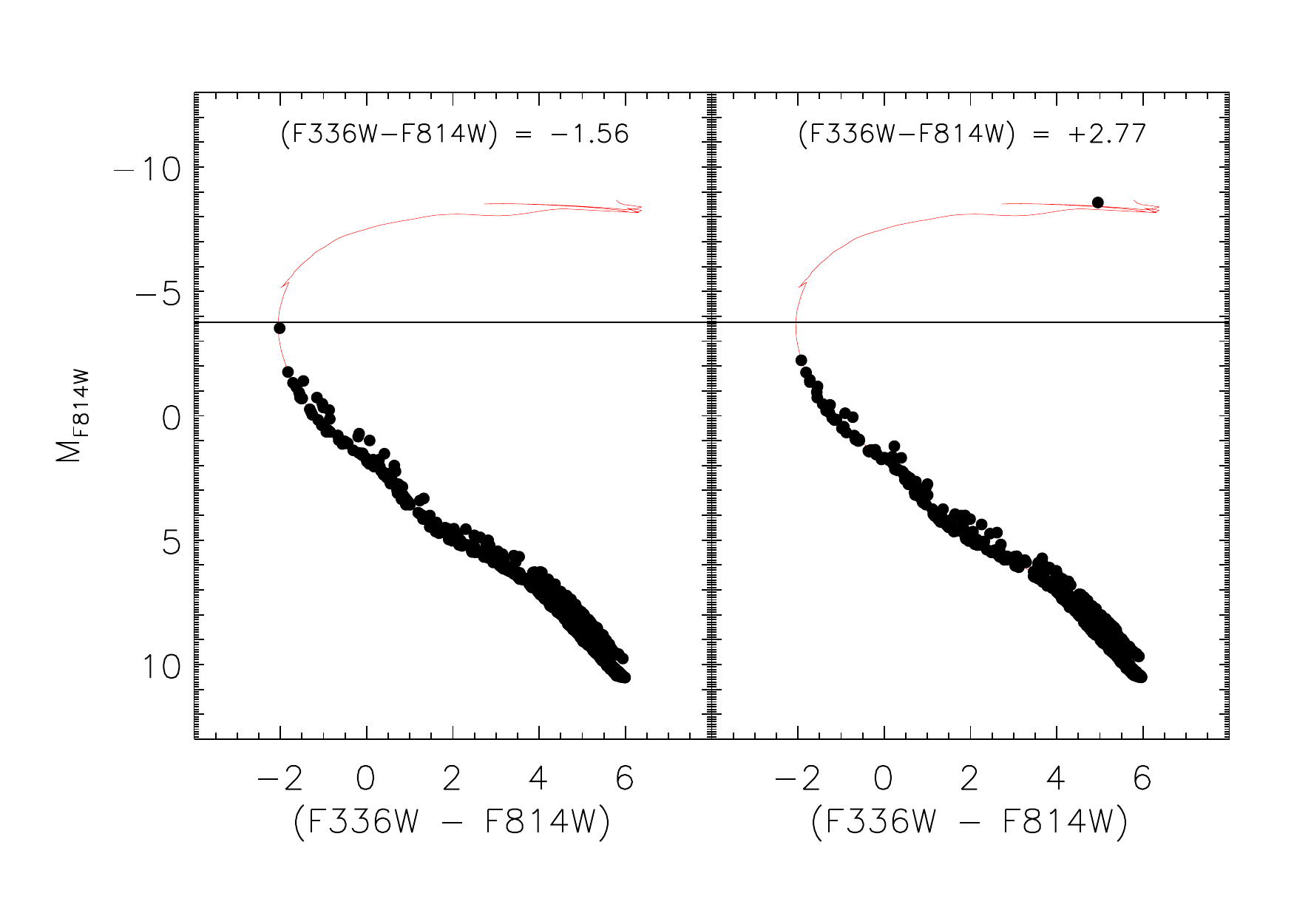}
}
\caption{
Simulated color-magnitude diagrams for two 10 Myr old, $10^3$ $M_{\odot}$ clusters at
solar metallicity.  Due to stochastic sampling of the initial mass function, the cluster in the
left panel does not contain any evolved stars, while the cluster on the right has
one evolved red giant. Integrated colors are included for each panel, along with 
a 10 Myr isochrone, and a horizontal line showing the magnitude of the main sequence turnoff.
\label{fig:example}}
\end{figure}

\begin{figure*}[!htbp]
\centerline{
\includegraphics[width=5in]{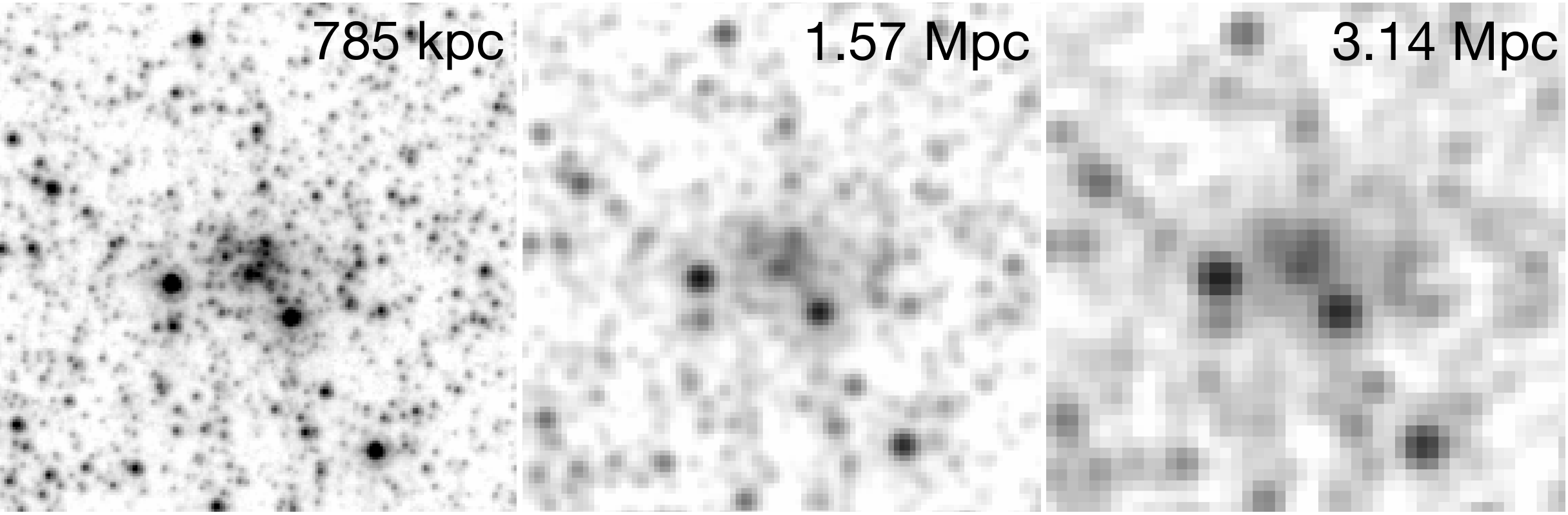}
}
\caption{
Examples of the variation in cluster resolution with distance.  The left panel shows the M31 cluster PC1017, at a distance of 0.785 Mpc \citep{Johnson12}.  The middle and right panels show image simulations of this cluster at roughly twice and four times the distance of M31 (1.57 and 3.14 Mpc).  The image is in the F814W filter, and has been smoothed and rebinned to simulate the two further distances.
\label{fig:resolution}}
\end{figure*}

Figure~\ref{fig:example} shows two realizations of a 10 Myr old cluster with a mass of $10^3$ $M_{\odot}$, generated using the program {\tt Fake} \citep{Dolphin02}, which is described in detail in \S\ref{sec:models}.  Both of these clusters are generated assuming the same underlying isochrone, total mass, and initial mass function.  The thickness of the main sequence is due to binarity, assuming a binary fraction of 0.35 and a \citet{Salpeter55} slope for the binary mass distribution.  In the realization in the left panel, the sampled population does not include any rapidly evolving post-main sequence stars.  The integrated (F336W$-$F814W) color is $-1.56$, and is dominated by the massive blue stars at the top of the main sequence.  In the right panel, however, the cluster (again, with the same mass and age) happens to have one very massive, bright red giant, due to stochastic sampling of the initial mass function. This bright evolved star now dominates the total integrated color of this cluster, which is now $+2.77$ -- a difference of more than four magnitudes.  A 10 Myr old cluster could therefore potentially have the same integrated color as a 10 Gyr old cluster due to one evolved giant star.  Although the effect is smaller, one might also expect stochastic variations in the blue flux due to variations in the sparsely populated upper main sequence.  This can also be seen in Figure~\ref{fig:example}, where there is a 1 magnitude difference in the luminosity of the brightest main sequence star between the two cluster realizations.  These effects make age and mass determinations based on integrated color extremely uncertain.

While Figure~\ref{fig:example} shows large differences in the two cluster realizations brightward of $M_{F814W}$ = $-$2, faintward of this limit the two clusters are populated quite similarly.  This similarity suggests that if the single red giant could be excluded, the color and luminosity of the rest of the stellar population would be better behaved, and would share the same blue colors expected for a young cluster.  This could in practice be done if the brightest one or several stars that correspond to the red giant and supergiant phases are resolved.

With HST's high resolution imaging, there are a number of large nearby galaxies whose clusters are partially resolved into stars.  Figure~\ref{fig:resolution} demonstrates the variation in cluster resolution with distance.  The left panel shows the M31 cluster PC1017, at a distance of 0.78 Mpc \citep{Johnson12}.  The middle and right panels show image simulations of this cluster at roughly twice and four times the distance of M31 (1.57 and 3.14 Mpc).  Even at these distances, the cluster's brightest stars can be resolved (in particular, the red evolved stars), while the majority of the more numerous main sequence stars remain unresolved.  These fainter stars are blended together and fall below the detection limit, particularly in the center of the cluster, where the crowding and background level are high.  Many older clusters at larger distances may also fall in this regime, because evaporation and cluster dissolution lead them to have lower surface densities, allowing individual bright stars to be detected \citep{Gieles11}.

\section{Method}  \label{sec:method}

\subsection{Simulated Clusters} \label{sec:models}

\begin{figure*}[!htbp]
\centerline{
\includegraphics[width=0.7\textwidth]{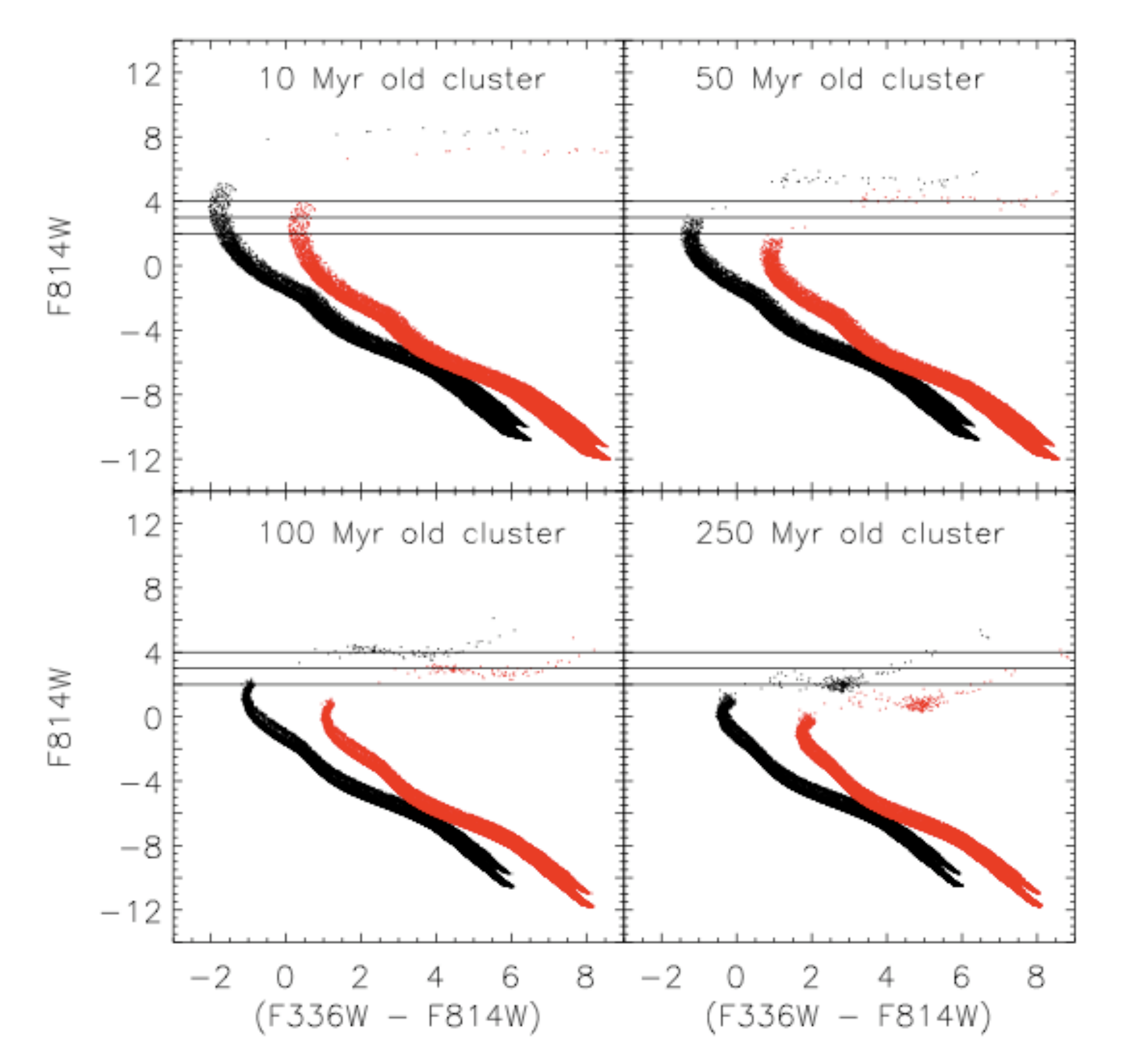}
}
\caption{Color-magnitude diagrams in two of the PHAT survey filters for solar metallicity clusters at four fiducial ages.  Horizontal lines indicate three possible values for $M_{\rm lim}$:  $-$2.0, $-$3.0, and $-$4.0.  Stars fainter than these lines are considered part of the unresolved component.  The black clusters have no extinction, while the red clusters have $A_{V}$ = 2.0 mag.
\label{fig:CMDcutoffs}}
\end{figure*}

To demonstrate the potential of analyzing the unresolved flux in a partially resolved cluster (rather than its fully integrated
counterpart), we start by building a set of unresolved light models.  This includes choosing a value for the magnitude cutoff, $M_{\rm lim}$, which is described in \S\ref{sec:cutoff}.  Then, based on this set of models, we show our method using simulated clusters and an M31 test cluster from the PHAT data set.

We utilized a variety of simulated clusters to test the degree to which stochastic effects are 
suppressed by restricting the light to the unresolved portion.  We generated CMDs for simulated clusters using the program {\tt Fake}, which is part of the CMD analysis suite MATCH (see \citealt{Dolphin02}).  This program generates CMDs for synthetic stellar populations using theoretical isochrones.  We used
the Padova models \citep{Marigo08} for the isochrone set, with updated AGB tracks from \citealt{Girardi10}, along with a Salpeter initial mass function \citep{Salpeter55}.  The Salpeter IMF produces more low-mass stars and a higher mass-to-light ratio than the Kroupa IMF \citep{Kroupa01}.  The simulations were done using HST filters, since we plan to apply this technique to M31 clusters observed with HST.  We chose to study clusters at four fiducial masses: $M_{cl}$ / $M_{\odot}$ = $10^3$, $10^4$, $10^5$, and $10^6$.  This range is representative of the various masses of clusters we expect to
be able to detect in nearby galaxies with HST.  At the lower mass end, the clusters experience a high level of stochasticity, while at the high mass end, clusters fully populate their isochrones, and suffer few color and luminosity variations due to stochasticity.  The simulated clusters' ages $t$ ranged from log($t/{\rm yr}$) = 7 to log($t/{\rm yr}$) = 10, in increments of $\Delta$ log($t/{\rm yr}$) = 0.3 $-$ 0.4.  Thirty clusters at each age and mass were produced at solar metallicity, and another thirty at subsolar metallicity (Z = 0.1 $Z_{\odot}$).  Solar metallicity clusters are mainly used throughout this paper since this is appropriate for young clusters in nearby large spirals like M31.  

{\tt Fake} generates clusters by randomly drawing masses from the initial mass function until the desired cluster mass is reached.  However, if a massive star is drawn which will put the cluster over the given limit, this star is discarded in favor of lower mass stars.  This approach leads to a bias in the resulting cluster's mass function, which will include more low mass main sequence stars.  This is a known effect of the cluster simulation, and should be kept in mind when analyzing resulting cluster fluxes and colors.  However, we are not studying the clusters' mass functions, and this does not affect the self-consistent analysis on synthetic models that we conduct in this paper.  Alternate methods exist to sample the cluster mass function, including those found in \citep{Apellaniz09, Weidner06, Weidner04, Popescu09}, some of which produce better sampling of the mass function of very low mass clusters.  However, the differences between our method and a proper filling of the cluster mass function are small for the mass range we are considering ($M_{cl}$ $\gtrsim$ $10^3$ $M_{\odot}$).

\subsection{Selection of Magnitude Cutoff}\label{sec:cutoff}

To study the broad behavior of partially resolved clusters, we chose a simple magnitude cutoff $M_{\rm lim}$ that defines the threshold below which the 
cluster stars are treated as unresolved.  This threshold does not necessarily have to be assigned to the actual magnitude limit of the data, and should be far from the completeness limit.  Such an approach can compensate for the strong radial variation in the detection limit, which tends to be significantly brighter in a cluster's crowded inner regions.  Ideally, we would like our value of $M_{\rm lim}$ to be at a level that excludes all the luminous evolved stars, and that leaves stars up through the main sequence turnoff.  If our $M_{\rm lim}$ is too faint, the main sequence turnoff will not be included, thus excluding the source of the age information.  If our imposed $M_{\rm lim}$ is too bright, we will include more post-main sequence stars, which will not be as effective in reducing stochasticity.  In this section we examine the best choice of $M_{\rm lim}$ for the PHAT data set.  A similar exercise should be done when applying this method to other data sets.

\begin{figure*}[!htbp]
\centerline{
\includegraphics[width=0.7\textwidth]{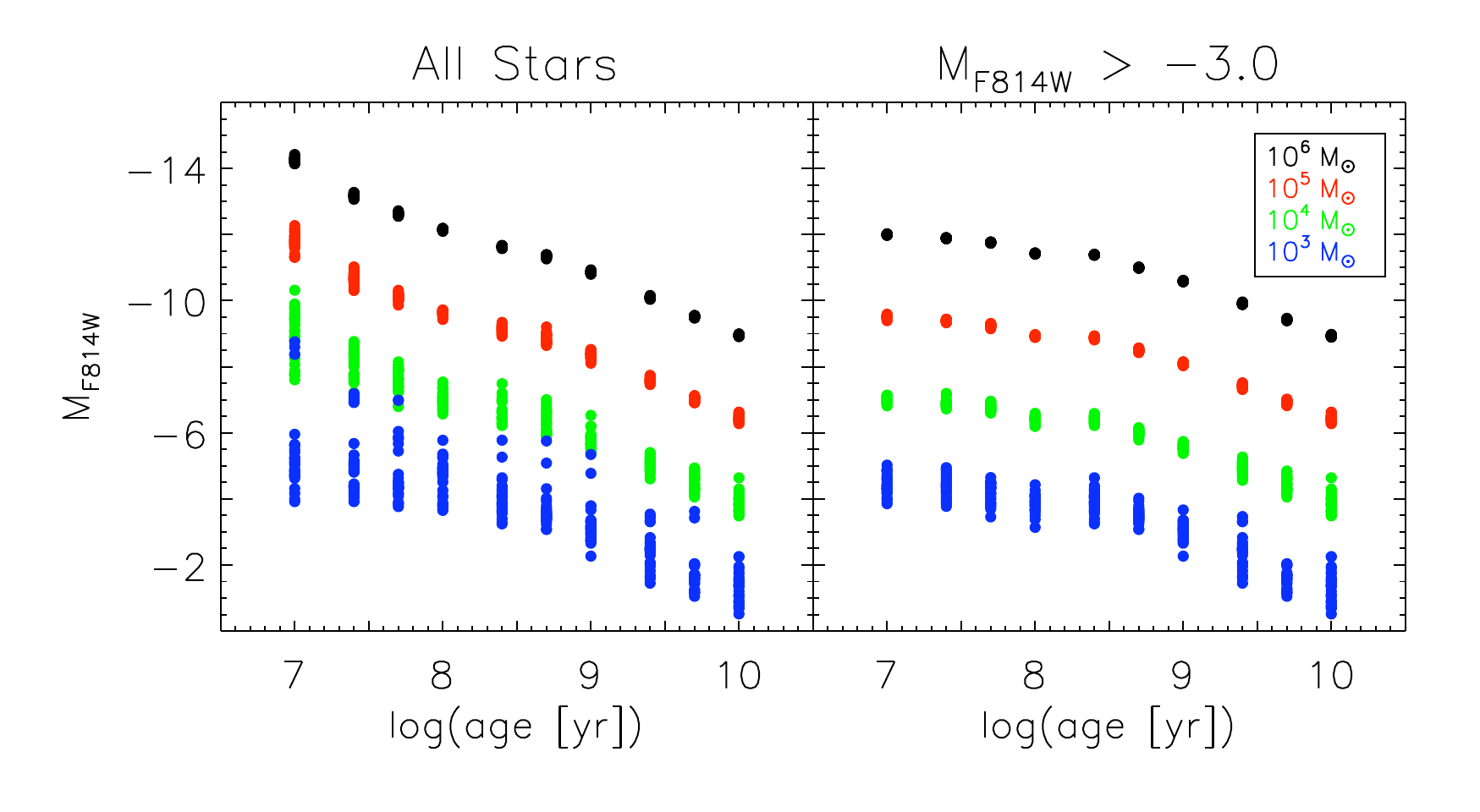}
}
\caption{
Absolute magnitudes in the F814W filter as a function of age for simulated solar metallicity clusters at four fiducial masses.  
The left panel shows the total integrated flux, while the right panel shows only the 
unresolved flux contained in stars fainter than the magnitude cutoff at $M_{\rm lim}$ = $-$3.
30 clusters are plotted for each age.  The clusters in the right panel exhibit much less spread 
in magnitude at a given age and mass, since the most luminous, rapidly evolving stars are excluded.
\label{fig:fluxcompare}}
\end{figure*}

Figure~\ref{fig:CMDcutoffs} shows the CMDs of simulated clusters at four fiducial ages, along with three values of $M_{\rm lim}$:  $-$2.0, $-$3.0, and $-$4.0.  
These values of $M_{\rm lim}$ were chosen to be above the detection thresholds of individual stars in M31.
The black points are for stars in a cluster with no extinction, and the red points are for $A_{V}$ = 2.0 mag (assuming $R_{V}$ = 3.1 and a \citet{Cardelli89} extinction law, which we use throughout this paper).  The number of stars being counted as part of the unresolved light clearly depends upon the age, extinction, and value of $M_{\rm lim}$.  Clusters younger than 25 Myr have all evolved stars cut off, regardless of $M_{\rm lim}$ and extinction.  For heavily extincted 50 Myr old clusters, some 
evolved stars remain below the less stringent values of $M_{\rm lim}$, and are therefore not removed from the unresolved component.  For clusters 
with little to no extinction, the age at which evolved stars begin to fall below the brightest $M_{\rm lim}$ is 100 Myr.  We therefore expect that this method will offer the most improvements when estimating the ages of younger clusters, which have F814W turnoff magnitudes of $-$2 and $-$3 for ages of 35 Myr and 18 Myr, respectively  \citep{Girardi10}.  Since younger clusters also suffer more from stochastic effects, we therefore focus on minimizing errors for younger clusters.  The tests described in  \S\ref{sec:tests} show reduced errors for $M_{\rm lim}$ = $-$2 and $-$3.  For our analysis, since $M_{\rm lim}$ = $-$2 and $-$3 are both effective at reducing stochastic effects in the young clusters, we choose to use the brighter $M_{\rm lim}$ = $-$3 to ensure that all the stars brighter than this will be well resolved.

The clusters in the PHAT survey are an excellent application for this method since they are observed in six filters with HST, and the completeness is high even at $M_{814}$ = 0.  At larger distances, this method can be used in galaxies that meet this resolution limit, but do no have enough resolved cluster stars to do isochrone fitting.  It is also important to note that the cluster photometry needs to be complete at or brighter than $M_{\rm lim}$.  This condition may be difficult to achieve for very dense clusters.

\subsection{Simulating the Unresolved Light}\label{sec:unres}

Figure~\ref{fig:fluxcompare} shows the relationship between a cluster's absolute magnitude and mass, for the integrated F814W magnitude of both the entire cluster (left panel) and the unresolved light below the magnitude threshold of $M_{\rm lim}$ = $-$3  (right panel).  For a given cluster mass, the variance in absolute magnitude depends on the cluster's age and number of evolved stars.  For the youngest, least massive clusters, the spread is almost five magnitudes, while the unresolved light spans less than 1.2 magnitudes.  This large magnitude dispersion is consistent with previous work on stochastic fluctuations \citep[e.g.,][]{Santos97, Piskunov09, Popescu10a}.  If a young low mass cluster contains only one or two bright, evolved giants or supergiants, that cluster's brightness can increase by several magnitudes, and it may appear as a more massive cluster.  In contrast, the clusters with no evolved stars have much fainter magnitudes.  This causes the bimodal distribution seen for the young low mass clusters, as previously pointed out by \citet{Chiosi88, Lancon00, Cervino03, Popescu10a, Popescu10b, Fouesneau10}.

\begin{figure*}[!htbp]
\centerline{
\includegraphics[width=0.7\textwidth]{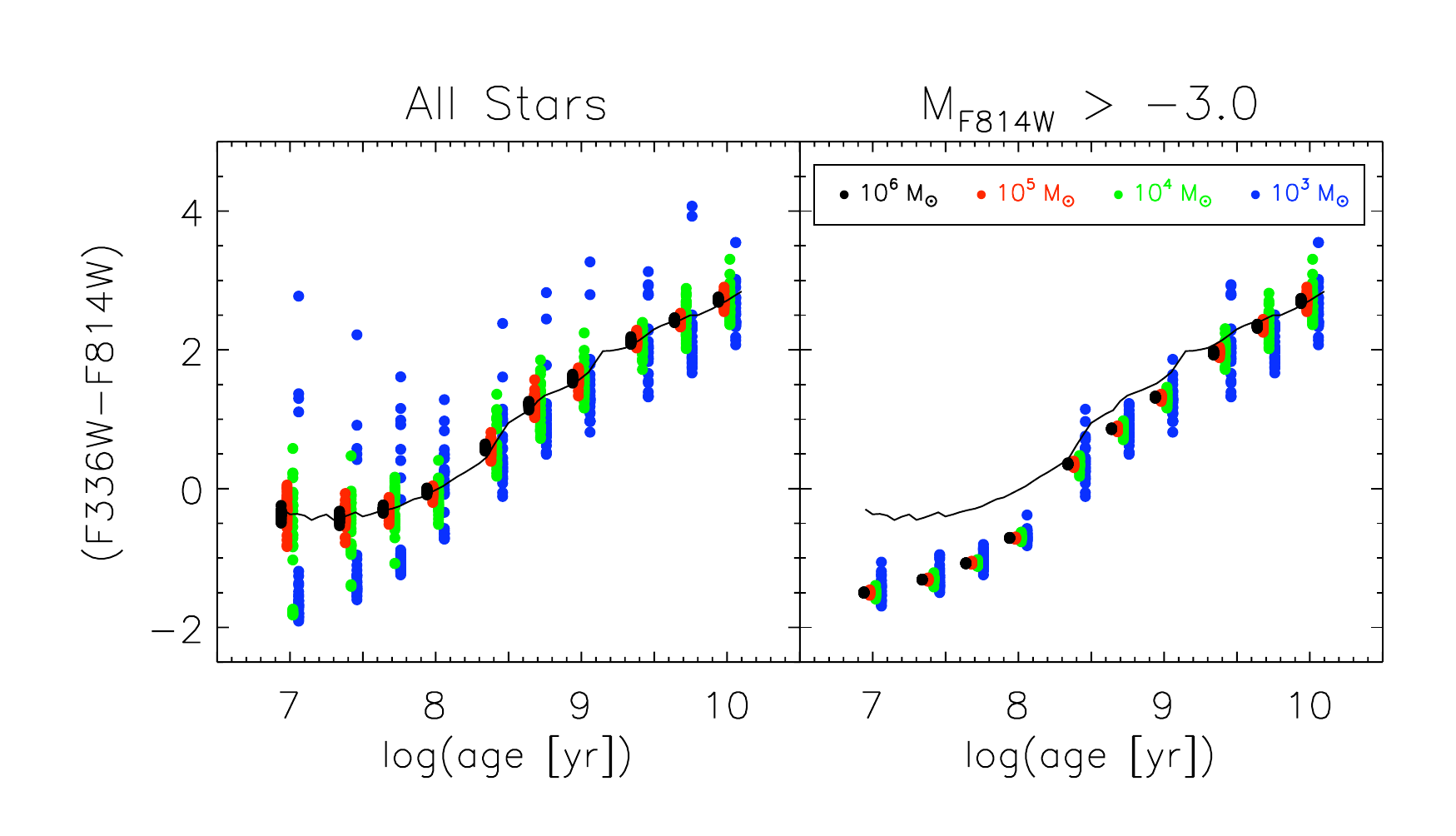}
}
\caption{
Colors as a function of age for simulated solar metallicity clusters at four fiducial masses.  
The left panel shows the total integrated color,
while the right panel shows only the unresolved color (excluding stars with F814W magnitudes 
brighter than $M_{\rm lim}$ = $-$3).  30 clusters are plotted for each age.  The break in the 
right panel at log($t$) $\ge$ 8.4 is due to $M_{\rm lim}$ being above the main sequence turnoff for older 
clusters, such that some evolved stars are included in the unresolved component.  
For clarity, all the $10^6$ $M_{\odot}$ clusters were shifted to the left by 0.06 dex,
the $10^5$ $M_{\odot}$ clusters were shifted to the left by 0.02 dex, the $10^4$ $M_{\odot}$ clusters 
were shifted to the right by 0.02 dex, and the $10^3$ $M_{\odot}$ clusters were shifted to the right by 
0.06 dex.  The black curve shows the colors predicted from continuously populated Padova SSP models \citep{Marigo08, Girardi10}.
\label{fig:colorcompare}}
\end{figure*}

The stochastic effects visible in Figures~\ref{fig:example} and~\ref{fig:fluxcompare} make age determinations using integrated colors challenging.  Theoretically, it is possible to obtain much better results by looking only at colors derived from the unresolved light.  To test this, in Figure~\ref{fig:colorcompare} we compare the colors derived from either the integrated light (left) or the unresolved light (right) to the simulated clusters' input ages, again restricting the unresolved light to stars fainter than $M_{\rm lim}$ = $-$3 in the F814W filter.

Figure~\ref{fig:colorcompare} shows that the spread in colors becomes small (less than 0.5 magnitudes) as the cluster mass approaches $10^6$ $M_{\odot}$ and the cluster's mass function becomes less subject to stochastic sampling.  Even for clusters of $10^5$ $M_{\odot}$, the colors of the youngest clusters span a range of 0.9 mag, which can lead to significant errors (more than 100 Myr) when deriving ages from these colors.  These uncertainties can cause significant problems in studies of more distant galaxies, for which clusters like these are the only clusters detected, and for which no individual stars can be resolved.

At even lower masses ($\lesssim$ $10^4$ $M_{\odot}$), stochasticity has an even larger effect, for the younger clusters especially.  The youngest clusters exhibit a range in colors from (F336W$-$F814W) = $-$2 to $+$3, a spread of five magnitudes.  At red colors, the $10^3$ $M_{\odot}$ clusters have only one or two red giants or supergiants that are responsible for biasing the overall color.  The bluer clusters have no red (super)giants, such that their light is dominated by the O and B stars on the main sequence.  This again causes the bimodal distribution in color seen for the young low mass clusters.

In contrast, when only the unresolved light is considered (right panel of Figure~\ref{fig:colorcompare}), the variation in color decreases to less than one magnitude for clusters with ages less than 100 Myr, even for the least massive clusters.  Thus, using only unresolved light reduces variation in color by four magnitudes.  For older clusters, the variation in color for the
least massive clusters increases to almost two magnitudes, since not as many of the red evolved stars' light are excluded.  However, in practice, old low mass clusters are rare since
most have already dispersed.  Ignoring the older ($>$ 100 Myr), least massive clusters, the color variations are on the order of one magnitude or less for all clusters of the same age and mass when only their unresolved light is considered.  

In addition to much smaller dispersion, the unresolved colors show several differences when compared to the traditional simple stellar population (SSP) models (shown as the black curve).  The unresolved colors are in general bluer than the colors given by the SSP models, since not all of the evolved, red (super)giant stars' light is included.  The unresolved colors also show a smooth variation for ages less than log($t/{\rm yr}$) = 8, while there is a modest discontinuity in unresolved colors seen between log($t/{\rm yr}$) = 8 and 8.4.  This effect is a result of where $M_{\rm lim}$ is in relation to the main sequence turnoff (see Figure~\ref{fig:CMDcutoffs}) and the age step size used.  For clusters younger than log($t/{\rm yr}$) = 8.4, the unresolved component excludes all evolved stars, while for older clusters, the choice of $M_{\rm lim}$ leaves some evolved stars in the unresolved component.  The colors of the older clusters, therefore, match more closely with those predicted by the SSP models.

Similar tests conducted at a metallicity of Z = 0.1 $Z_{\odot}$ yielded qualitatively similar results to the solar metallicity case presented above.

\section{Tests on Simulated Clusters}   \label{sec:tests}

In practice, determining a cluster's properties is often done through a simultaneous derivation of age, mass, and extinction determined using $\chi^2$ minimization compared to a set of fiducial models,  \citep[e.g.,][]{Pasquali03, Hancock08}.  The value of $\chi_{k}^2$ for comparing data to a set of $k$ model parameters is given by

\begin{equation} \chi^2_{k}(M) = \sum_{i=1}^{N} \frac{(D_{i} - M \times \theta_{ik})^2} {\sigma_{i}^2} \label{eqn:chisq}
\end{equation}
where $D_{i}$ is the data flux in the $i$th filter, $\theta_{ik}$ is the $k$th model flux in the $i$th filter, $\sigma_{i}$ is the variance in the $i$th filter, M is the mass, and the summation is over the available filters.  The parameters of the model $k$ are then varied until $\chi_{k}^2$ is at a global minimum. 

A $\chi^2$ minimization was used to simultaneously recover the age and extinction for a set of test clusters to a set of model clusters, while optimizing for the mass, using only their unresolved flux.  The test clusters were drawn from the clusters used in the analysis in \S\ref{sec:unres}, with additional clusters between the ages 10 Myr and 100 Myr added to better probe the rapidly evolving stellar populations of young clusters.  Thirty clusters at each age were chosen as test clusters, for a total of 540 test clusters.  The test clusters' masses ranged from $M_{cl}$ / $M_{\odot}$ = $10^3$ - $10^6$.  We looked at four HST filters that are being used in the PHAT survey:  F336W, F475W, F814W, and F160W.  Each test cluster was then given a random value of foreground extinction between $A_{V}$ = 0 to $A_{V}$ = 3.0 mag; no differential extinction was included for this initial test.  A small amount of random Gaussian noise (${\sigma}$ = 0.05 mag) was also added to these clusters to simulate observational errors, which we assume are Gaussian.  This noise value was used for the variance, $\sigma_{i}$, through the conversion $\sigma_{flux} = flux \times (1 - 10^{-0.4  \: \sigma_{mag}})$.

The set of clusters that were used as models were taken from the most massive clusters at each age.  These model clusters were massive enough for their main sequence to be considered fully populated ($> 10^5$ $M_{\odot}$).  Altogether, we used model clusters at eighteen different ages from log($t/{\rm yr}$) = 7 to 10.

\begin{figure*}[!htbp]
\centerline{
\includegraphics[width=6in]{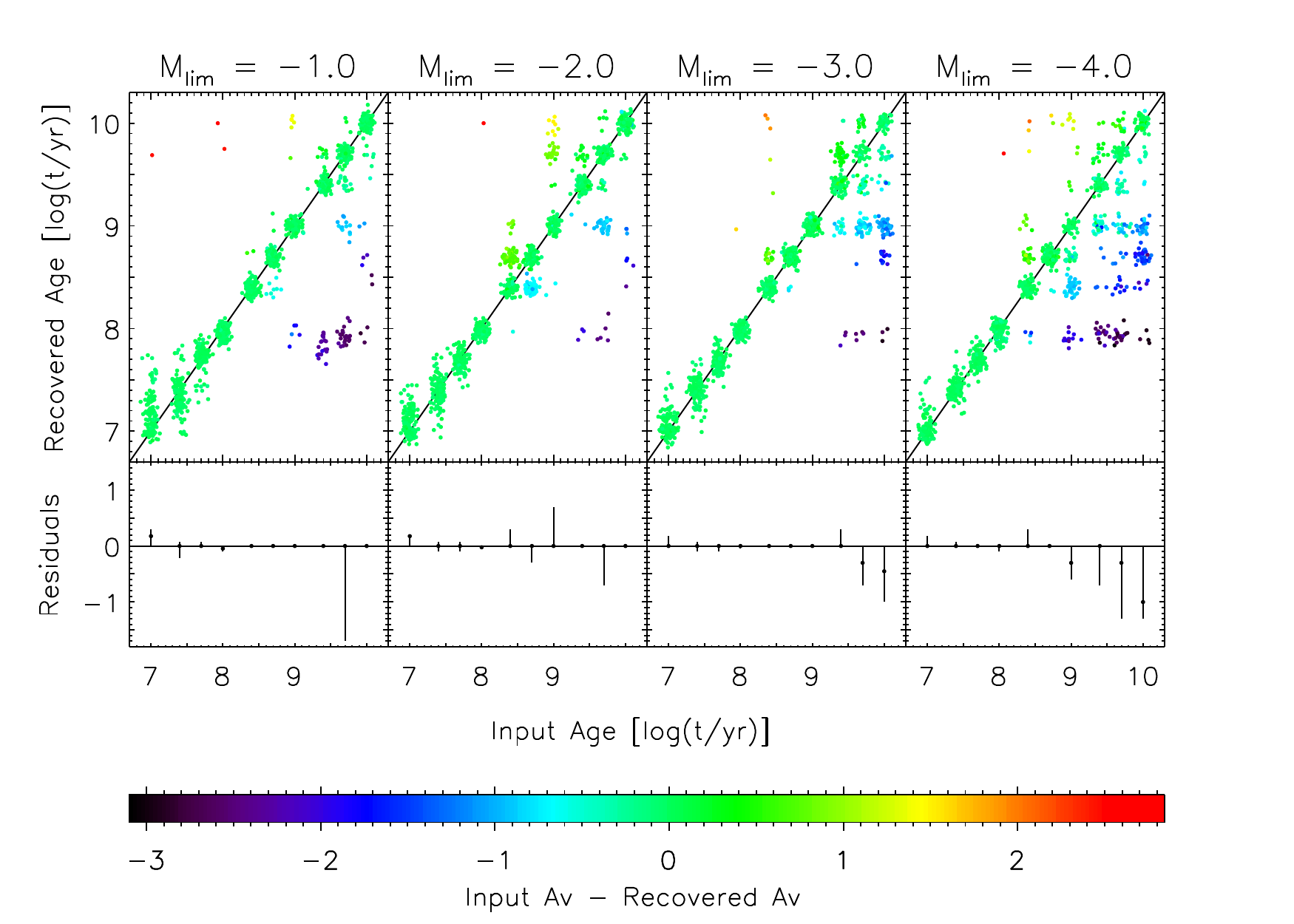}
}
\caption{Recovered vs input age of the simulated test clusters, color coded by the difference in recovered vs input extinction, for four values of $M_{\rm lim}$ in the F814W filter.  To clearly see the distribution of recovered values, Gaussian noise with a dispersion of $\sigma$ = 0.05 dex in each direction was added in the upper panels.  The bottom panels show the median for the residuals in log($t$) for all clusters at each input age, along with the 16$^{\textrm{th}}$ and 84$^{\textrm{th}}$ percentile of the residuals.
\label{fig:resultsunres}}
\end{figure*}

The models used were normalized as the unresolved flux per unit mass and, assuming that light scales as mass, the test clusters' unresolved fluxes were scaled to match the model clusters' fluxes.  Therefore, mass is a parameter solved for during the minimization by calculating the scale factor between the fluxes of the test clusters and the model clusters.  Extinction was recovered by including these same models at 50 values of extinction between $A_{V}$ = 0 to 3.5 mag.

Figure~\ref{fig:resultsunres} shows the recovered ages for four different values of $M_{\rm lim}$, color coded by the difference in recovered and input extinction.
For $M_{\rm lim}$ = $-$2, the bias in recovered ages is close to 0, and the average error is 0.09 dex.  The average error in mass is 0.05 dex, and there is a slight bias toward underestimating the mass.  The average error in $A_{V}$ is 0.12 mag.  The errors were calculated by taking the mean of the absolute differences in input $-$ output.  These values represent a lower limit, since the calculated errors are in many cases dominated by the grid size.  54\% of the clusters' ages are recovered to better than the resolution of the grid of models, and 96\% are recovered within 0.5 dex.  $M_{\rm lim}$ =  $-$3 has similar errors, while $M_{\rm lim}$ = $-$1 has a slightly larger error in age.  $M_{\rm lim}$ = $-$4 does not offer as large an improvement, and shows a large number of outliers.  With this brighter limit, evolved stars are still included in the unresolved light at most ages, so the stochastic effects are not as reduced.  For fainter magnitude limits ($M_{\rm lim}$ $\ge$ $-$1), the age sensitivity is reduced since all of the evolved stars and some of the bright main sequence stars are being removed  for clusters younger than 50 Myr with low extinction, as seen in Figure~\ref{fig:CMDcutoffs}.  For the least massive clusters (\textless $10^3$ $M_{\odot}$), the average error in age is 0.17 dex.  The feature seen at a recovered age of log($t/{\rm yr}$) = 8 consists mostly of low mass clusters that are not well recovered.  This occurs because, due to the shape of the continuous SSP models, when the minimization routine searches for the closest match to the models, certain age models are more attractive to a wider variety of clusters, which was was pointed out in Section 4.3 of \citealt{Fouesneau10}.

\begin{figure*}[!htbp]
\centering
\mbox{\subfigure{\includegraphics[width=0.75\textwidth]{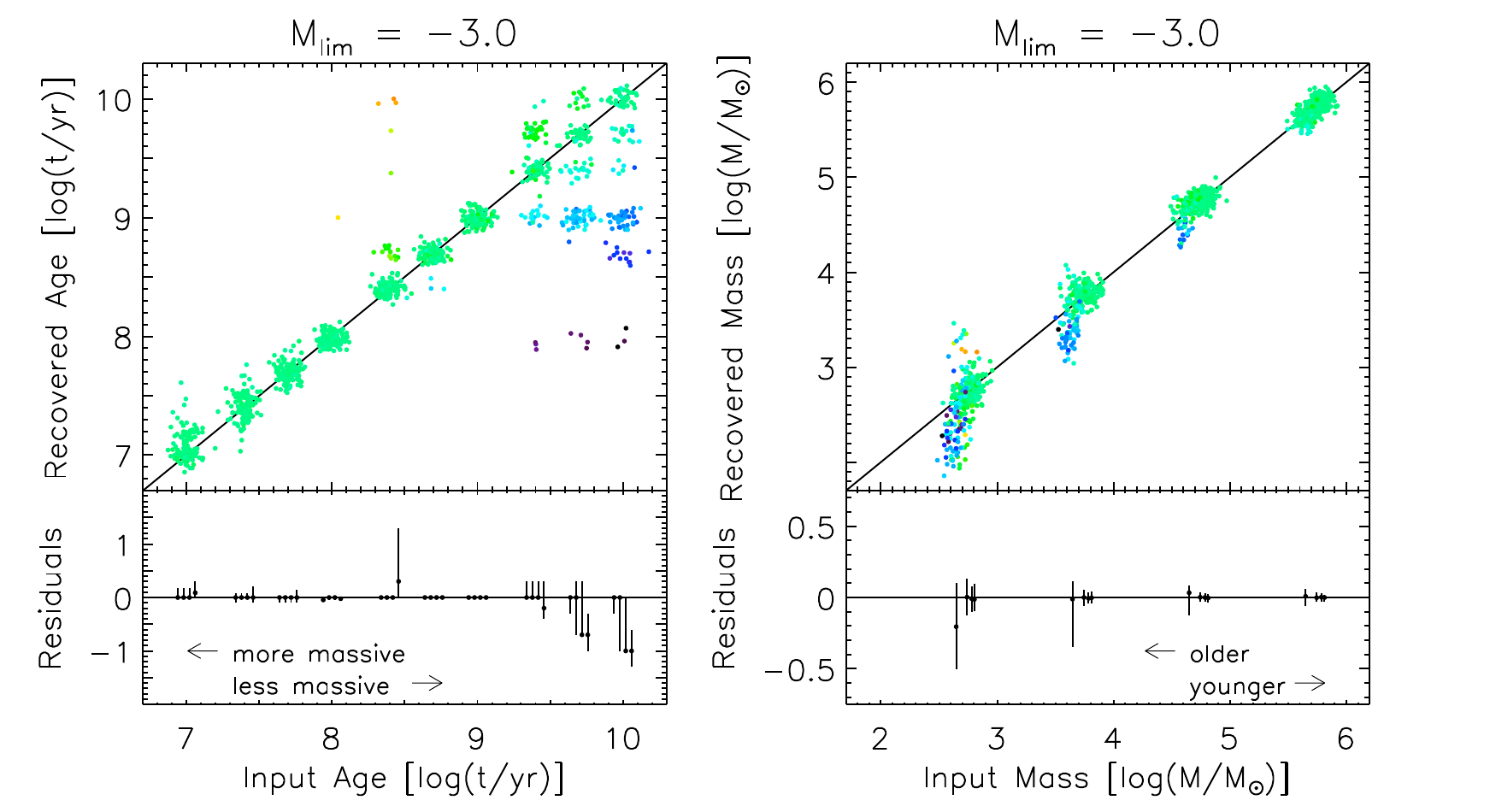}}}
\\
\mbox{\subfigure{\includegraphics[width=0.75\textwidth]{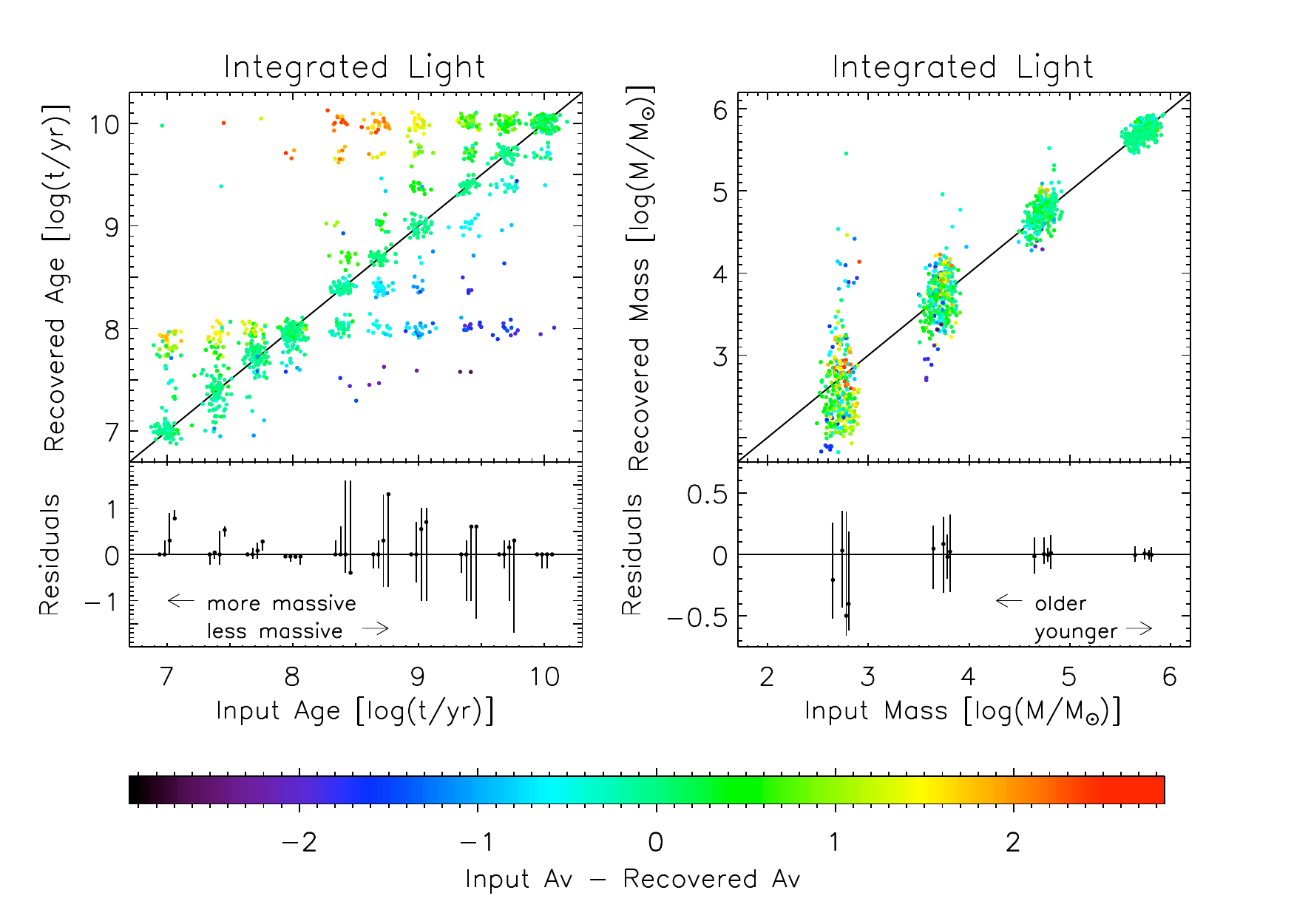}}}
\caption{Recovered vs input age (left plots) and mass (right plots) of the simulated test clusters, color coded by the variation in recovered extinction.  The top row shows the results from using unresolved light with $M_{\rm lim}$ = $-$3, while the bottom row shows the results from using integrated light.  We added Gaussian noise with a dispersion of $\sigma$ = 0.05 dex in each direction for clarity.  The bottom panels show the median (dot) along with the 16$^{\textrm{th}}$ and 84$^{\textrm{th}}$ percentile (bar) values of the residuals in log($t$) and log($M$), respectively.  At each age time step, the age residuals are shown for each input mass, where the more massive clusters' residuals are shifted slightly to the left and the less massive clusters are shifted slightly to the right.  The mass residuals are shown for four age groups (limits of 7.0, 7.5, 8.0, 8.5, 9.0), where the older clusters' residuals are shifted slightly to the left and the younger clusters' residuals are shifted slightly to the right.}
\label{fig:resultsint}
\end{figure*}

Figure~\ref{fig:resultsint} compares the results of the $\chi^2$ minimization for $M_{\rm lim}$ = $-$3 (top panel), with the results for traditional fitting to integrated light models (bottom panel).  The left plots show recovered ages and the right plots are the recovered masses, color coded by the difference in recovered and input extinction.  In addition to the feature seen at a recovered age of log($t/{\rm yr}$) = 8, explained above, there is an additional feature at log($t/{\rm yr}$) = 10.  This feature results from the finite range of model properties.  For clusters that are redder than the age-extinction range covered by the models, they are always fit to the oldest age in the models.  This effect can also be seen in \citealt{Fouesneau12}.  The large feature at a recovered age of log($t/{\rm yr}$) = 10 is not seen in the results from using the unresolved light (top panel).  The feature at log($t/{\rm yr}$) = 8 is also significantly reduced compared to the integrated light results.  Due to the monotonic variation in unresolved color, as seen in Figure~\ref{fig:colorcompare}, many of the sources of color degeneracies in integrated light are not present for the unresolved light.  This leads to greater accuracy in deriving the cluster ages and masses.

\begin{figure*}[!htbp]
\centerline{
\includegraphics[width=5in]{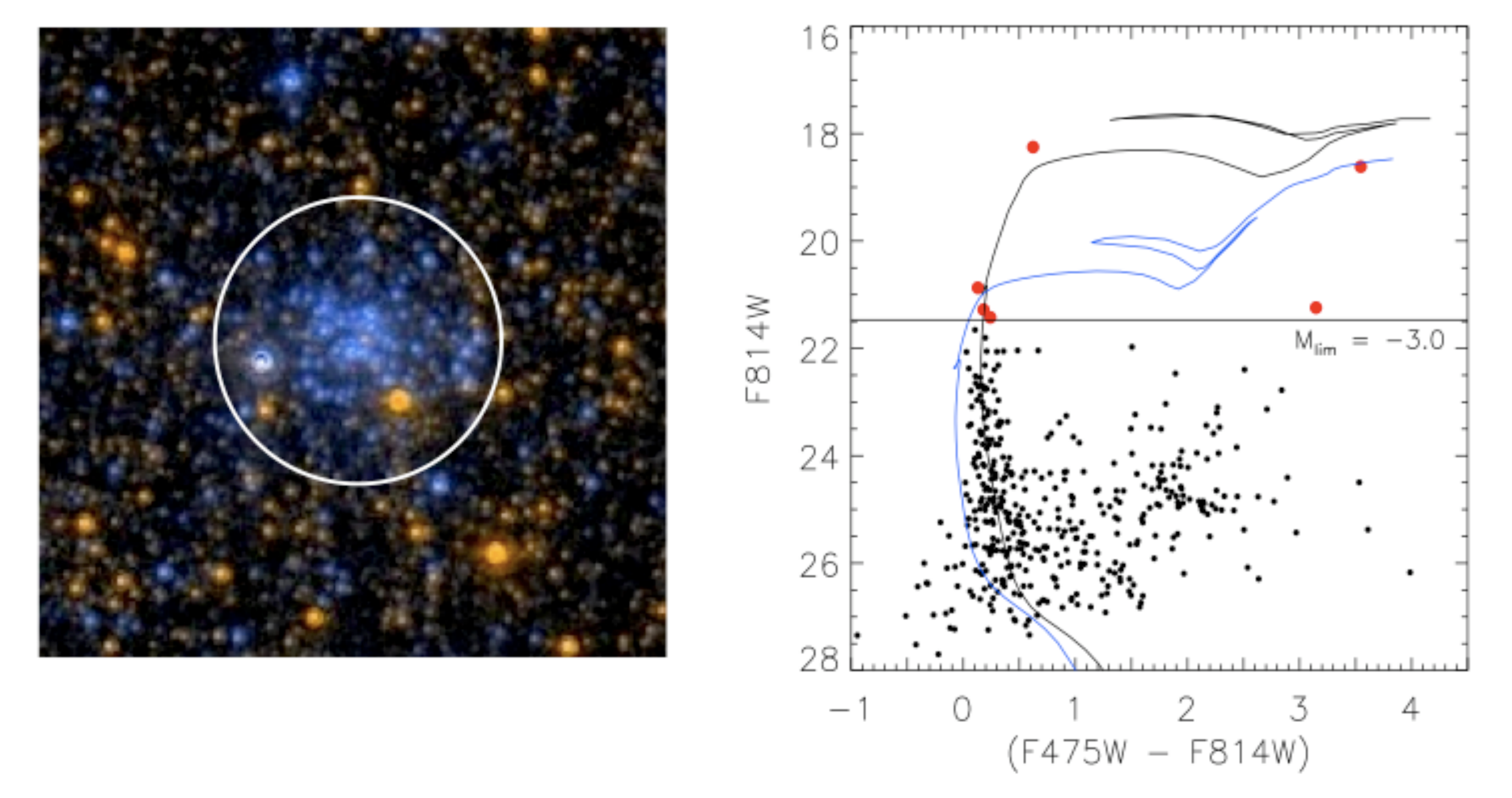}
}
\caption{Test cluster PC1017 in M31, as observed with the PHAT survey \citep{Johnson12}.  Left panel is the image shown with the photometric aperture.  Right panel shows the color-magnitude diagram for this same cluster.  The best-fit isochrone (20 Myr, $A_{V}$ = 0.9) from fitting the resolved stars is shown in black.  The stars excluded in the unresolved fitting are shown in red.  The unresolved method gives an age of 40 Myr and $A_{V}$ = 0.42.  The blue isochrone (80 Myr, $A_{V}$ = 0.28 mag) that was obtained using traditional integrated light fitting is a poor fit to the resolved stars.
\label{fig:real}}
\end{figure*}

For the fits to integrated light, there is a bias of 0.09 dex toward older ages, and the average error for all test clusters is 0.26 dex in age.  36\% of the clusters' ages are recovered to better than the resolution of the grid of models, and 83\% are recovered within 0.5 dex.  Some clusters' recovered ages can be as much as 3 dex away from their input age.  The mass estimates are more robust and have an average error of 0.19 dex.  However, significant scatter exists for the integrated light fits for the lower mass clusters, which violate the assumption that mass scales monotonically with luminosity (see Figure~\ref{fig:fluxcompare}), as we have implicitly assumed in our fitting process.  The average error for $A_{V}$ is 0.45 mag, and recovered values of $A_{V}$ are biased by 0.16 mag toward lower extinctions.  This extinction bias explains the age bias, since young, highly extincted clusters are being recovered as older, less extincted clusters.  When considering only clusters whose masses are $\le$ $10^3$ $M_{\odot}$, the average error in age is 0.51 dex, considerably higher than the error for low mass clusters when using unresolved light.

In this analysis we explored clusters with a range of extinction up to $A_{V}$ = 3.0 mag, which is appropriate for most M31 clusters.  If independent constraints on $A_{V}$ are available, the age-extinction degeneracy is lifted, and the errors will be reduced even further.  For example, for $M_{\rm lim}$  = $-$2, if  $A_{V}$ is known to within 1 magnitude, the error in age decreases from 0.09 to 0.03 dex.  Also, using additional filters would improve the accuracy of the age and mass determination \citep{Apellaniz09},  while providing greater constraints on extinction.  The metallicity of all of our clusters was fixed at solar, however this analysis could be extended to varying metallicities.  Allowing the metallicity to vary would cause further degeneracies with age and extinction.  A full characterization of errors and systematics is beyond the scope of this paper and is postponed to a further publication.

\section{Application to a Real Cluster}   \label{sec:application}

In this section we extend the $\chi^2$ fitting to an observed cluster from the PHAT cluster catalog \citep{Johnson12}, PC1017.  Figure~\ref{fig:real} shows the CMD for cluster PC1017 which we use as a test cluster to demonstrate our unresolved method.

The challenge of using a real cluster to test this method is that we do not know its intrinsic properties.  We chose this cluster as our test cluster because there are good independent estimates of its parameters.  \citet{Caldwell09} used spectroscopy to age-date this cluster as 33 Myr, with a factor of 2 uncertainty.  Its mass was found to be log($M_{cl}$ / $M_{\odot}$) = 3.88, and $A_{V}$ was 0.68 mag.  Stochasticity can however affect these spectroscopic estimates as well.  Another prediction given by the discrete models from \citealt{Fouesneau10} gives an age of 41 Myr, log($M_{cl}$ / $M_{\odot}$) = 3.9, and $A_{V}$ = 0.3 mag.  Additionally, since this cluster has many resolved stars, we were able to make determinations by isochrone matching the best fit age and extinction with the CMD analysis suite MATCH \citep{Dolphin02}.  This gives an age of 20 Myr (log($t/{\rm yr}$) = 7.3 +/- 0.1),  log($M_{cl}$) = 3.8 +/- 0.3, and an Av = 0.9 +/- 0.15 mag.

We measure the unresolved flux for the cluster using a three step process.  Due to large photometric errors in F160W, we use only F336W, F475W, and F814W.  For each of the six images, we sum the light within the photometric aperture (2.19") and within a background annulus (2.63" - 7.45").  Next, we identify all stars with F814W magnitudes brighter than $M_{\rm lim}$ = $-$3 within the two measurement regions.  This removes the flux from the six brightest stars in Figure~\ref{fig:real}.  We subtract the flux contribution from these bright stars from the cluster and background flux totals.  Finally, we subtract the remaining background flux (representing the summed light of field stars with F814W \textgreater $M_{\rm lim}$) and obtain our final unresolved fluxes for the cluster.  

The cluster's properties were determined through $\chi^2$ minimization using the unresolved flux values.  For PC1017, the unresolved light analysis gives a best fit age of 40 Myr, with log($M_{cl}$) = 3.62 $M_{\odot}$ and $A_{V}$ = 0.42 mag.  All models with $\chi^2$ $<$   1 $\pm$ (1 / $\sqrt{n}$) were considered good fits, where $n$ represents the degrees of freedom, 3 in this case.  From the distribution of good fits, the 68\% confidence interval gives an age range of 16 - 64 Myr, a mass range of log($M_{cl}$) = 3.56 - 3.67 $M_{\odot}$, and an extinction between 0.20 and 0.64 mag.  These values are consistent with both the spectroscopically determined properties, as well as those predicted by the discrete models.

We have also analyzed this cluster using traditional integrated light fitting.  This method gives a best fit age of 80 Myr, log($M_{cl}$) of 3.86 $M_{\odot}$, and $A_{V}$ of 0.28 mag.  The 68\% confidence intervals were 64 - 96 Myr, with a log($M_{cl}$) between 3.81 and 3.90 $M_{\odot}$, and an extinction between 0.17 and 0.39 mag.  Fitting the total integrated light therefore gives a derived age that is too old, due to the presence of the luminous red giant.  In contrast, the unresolved flux results in an age that is more consistent with the ages determined spectroscopically, by the discrete models, and from isochrone fitting.  

We find that the mass estimated by the unresolved method is lower by $\sim$ 0.2 dex compared to the other three methods, which is of the same order as the bias towards low masses found in \S\ref{sec:tests}.  However, the unresolved mass estimate is still within the error bars determined by the isochrone fitting as well.  The underestimate of the cluster's mass could indicate that the mass-to-light scaling needs to be recalibrated for use with unresolved light.

Though there are differences in the derived ages, masses, and extinctions for the different methods, the properties determined by the unresolved method agree well with those determined by the independent estimates.  The differences in extinction can mostly be attributed to the age-extinction degeneracy.  The age determined by isochrone fitting was the youngest, and it also had the highest extinction.  Alternately, this cluster could be slightly older, with a lower extinction, as estimated by the discrete and unresolved models.

\section{Discussion}   \label{sec:discussion}

\subsection{Applicability}\label{sec:applicability}
Analyzing the unresolved component of a cluster's light shows promise for reducing some of the stochastic effects associated with deriving the properties of star clusters.  This method is best suited for low mass clusters, where stochastic issues make normal integrated light methods fail, and where there are small numbers of bright stars that we wish to subtract.  The dispersion is still significant for clusters of $10^3$ $M_{\odot}$, but this method can produce more accurate results for these lower mass clusters than traditional integrated light fitting.  It can be used for clusters in galaxies from $\sim$ 1-3 Mpc, where full CMD fitting is not possible, but a small number of bright stars can be individually resolved.  The value of $M_{\rm lim}$ should be optimized for the targeted clusters' age range.  

\subsection{Best Choice of $M_{\rm lim}$}\label{sec:choice}
The choice of cutoff magnitude $M_{\rm lim}$ was based on trying to achieve stability in the unresolved component of the clusters' light, while preserving a strong correlation of flux with age.  The values of $M_{\rm lim}$ discussed in this paper were chosen to work well with the PHAT data set.  Several things should be kept in mind when trying to decide what value of $M_{\rm lim}$ should be used.  At distances greater than a few Mpc, even HST imaging will make it difficult to resolve stars down to moderately bright magnitudes (e.g., $M_{814}$ = $-$3).  However, in these cases, the method should be adjusted to use a brighter $M_{\rm lim}$ at the expense of limiting the age range one can probe.  A brighter cutoff would be better suited to younger and/or less extincted clusters, while a fainter cutoff would yield better results for older and/or more extincted clusters.

One difficulty associated with subtracting the resolved star fluxes is that completeness of resolved photometry can vary greatly as a function of radius within the cluster.  $M_{\rm lim}$ should be above the completeness limit of the data since accurate stellar photometry of the resolved stars is needed to subtract off their flux.  If $M_{\rm lim}$ is bright enough and the number of bright stars we wish to subtract is small enough, then resolving these few bright objects should be possible in most cases.  This will be the case for lower mass clusters, for which this method is most useful.  To optimize this method for a general use, $M_{\rm lim}$ could be a parameter that is solved for during the $\chi^2$ minimization as well.  This would allow the value of $M_{\rm lim}$ to be optimized for the age of each cluster, and to reduce stochastic effects for older clusters as well as younger clusters.

\subsection{Field/Foreground Contamination}\label{sec:contamination}
Another benefit to this method not previously discussed in that it minimizes bright field/foreground star contamination.  As long as the star in question is brighter than the chosen $M_{\rm lim}$, its flux is not included in the analysis.  Therefore, the determination of cluster membership does not affect the derived cluster's properties.

\subsection{Model Uncertainties}\label{sec:uncertainties}
Our analysis depends on the accuracy of the models used and the assumption that dust attenuation acts in a predictable way.  These factors may cause the actual uncertainties in derived properties to be higher.  One advantage of looking at the unresolved light is that this will be less sensitive to model uncertainties for massive post-main sequence stars, which are quite substantial.  One potential complication, however, is that the separation between main sequence and post-main sequence stars is not always as clean as the models predict \citep{Larsen11}, which again complicates the choice of $M_{\rm lim}$.  

\subsection{Future Work}\label{sec:future}
Eventually the best way to study a cluster is to combine the study of the unresolved and resolved portions of its light.  The resolved stars can be analyzed using isochrone fitting, and the unresolved color could be analyzed separately.  Then the results from these two methods can be compared to ensure consistency.

A more complete study of a variety of clusters is needed to further show the applicability of this method.  This would include optimizing the value of $M_{\rm lim}$ to be solved for during the fitting process, investigating the effects of crowding and blending, further comparisons with other age-dating methods, and extending this method out to larger distances.

\section{Conclusions}   \label{sec:conclusion}

We investigated the properties of partially resolved stellar clusters using simulations and found that stochastic variations in color can be greatly decreased when
considering only flux below a limiting magnitude $M_{\rm lim}$.  This unresolved light component utilizes the stability of the main sequence light along with the age information in the main sequence turnoff but eliminates the often stochastically sampled upper end of the stellar mass function.  By using only this unresolved component of the flux, we have shown that we can derive accurate age, mass, and extinction determinations with a variety of simulated clusters.  The improvements over traditional integrated light fitting are most evident for lower mass clusters where the effects of stochasticity are greatest.  This method was also applied to an M31 cluster, and results using the unresolved method were comparable to the properties determined spectroscopically, from discrete models, and isochrone fitting.

This new technique will be especially useful for lower mass clusters (less than a few $10^4$ $M_{\odot}$), crowded clusters, and clusters in nearby galaxies with only a few resolved stars.  This method can also be used as a sanity check for clusters whose age and mass determinations come from isochrone fitting.  It also allows for the potential of obtaining reliable property determination for clusters too far away to be analyzed with isochrone fitting methods.  In this situation, the unresolved technique provides greater accuracy while still utilizing continuous models, without needing spectroscopy or computationally intensive discrete models.

\acknowledgements

{The authors wish to acknowledge the collective efforts of the entire PHAT team in this project.  Also, the authors thank the anonymous referee for a prompt and useful report.  This research made extensive use of NASA's Astrophysics Data System Bibliographic Services.  Support for this work was provided by NASA through grant number HST-GO-12055 from the Space Telescope Science Institute, which is operated by AURA, Inc., under NASA contract NAS5-26555.  D.A.G. kindly acknowledges support by the German Research Foundation (DFG) through grant GO 1659/3-1.}  


\bibliographystyle{apj}


\end{document}